\documentclass[12pt,amsfonts]{article}    
\usepackage{amsfonts}

\def\half{\textstyle{\frac{1}{2}}}

\def\ra{\rightarrow}
\def\tint{{\textstyle\int}}

\def\b{\begin{eqnarray*}}  
\def\e{\end{eqnarray*}}    
\def\bn{\begin{eqnarray}}  
\def\en{\end{eqnarray}}   

\def\<{\langle}
\def\>{\rangle}

\def\{{\lbrace}
\def\}{\rbrace}

\def\half{\textstyle{\frac{1}{2}}}

\def\qb{\bar{q}}
\def\pb{\bar{p}}
\def\Pb{\bar{P}}
\def\Qb{\bar{Q}}
\def\Hb{\bar{H}}

\begin{document}

\title{Is Loop Quantum Gravity \\ a Physically Correct Quantization?}
 
\author{John R. Klauder\footnote{klauder@phys.ufl.edu} \\
Department of Physics and Department of Mathematics \\
University of Florida,   
Gainesville, FL 32611-8440} 
\date{ }
\bibliographystyle{unsrt}
\maketitle

\begin{abstract} Dirac's rule in which only special phase space variables should be promoted to operators in canonical quantization is applied to loop quantum gravity. For this theory, Dirac's rule is violated, and as a result loop quantum gravity fails the test to be a valid quantization. Indications are included on how to create and deal with valid versions of 
quantum gravity.\footnote{DOI: 10.4236/jhepgc.2020.61006}
\end{abstract} 
{\bf Key words:} quantum gravity, loop quantum gravity, Dirac's rule

\section{Dirac's Rule for Canonical Quantization} For a single degree of freedom, a momentum $p$ and a position $q$, where $-\infty<p, q<\infty$,
the Poisson bracket is $\{q,p\}=1$, and the Hamiltonian function is given by
$H(p,q)$. In addition, new variables\footnote{e.g., $\pb=p/q^2$ and $\qb=q^3/3$.} may also be used, say, $\pb$ and $\qb$, $\{\qb, \pb\}=1$, $-\infty<\pb, \qb<\infty$, and $\Hb(\pb,\qb)=H(p,q).$ 

For canonical quantization, we promote $p\ra  P$, 
$q\ra Q$, $[Q,P]=i\hbar$, along with $H(p,q)\ra
H(P,Q)$. In addition,  $\pb\ra\Pb$, $\qb\ra\Qb$, $[\Qb,\Pb]=i\hbar$, and $\Hb(
\pb,\qb)\ra \Hb(\Pb,\Qb)$, {\bf BUT}, $ \Hb(\Pb,\Qb)\neq H(P,Q)$. At most, only one such quantization
can be valid while all others lead to false quantizations.

Although the classical Hamiltonians can be equal the quantum Hamiltonians are different, and the question arises which is the physically correct  Hamiltonian operator. Dirac \cite{D} asserts that the proper choice of the quantum Hamiltonian is the one which has been promoted from Cartesian coordinates as classical variables.\footnote{In particular, in the mid-page of 114 Dirac wrote ``However, if the system does have a classical analogue, its connexion with classical mechanics is specially close and one can usually assume that
the Hamiltonian is the same function of the canonical coordinates and momenta in the quantum theory
as in the classical theory.$\dag$ "
Footnote ``$\dag$  This assumption is found in practice to be successful only when applied
with the dynamical coordinates and momenta referring to a Cartesian system of axes and not
to more general curvilinear coordinates."} Dirac
does not prove his rule, but Dirac's rule has recently been established
\cite{JK} leading to a flat space (Fubini-Study) metric given by $d\sigma(p,q)^2= A \,dp^2+A^{-1} \,dq^2$, where
 $A>0$ is a constant. Although we have focussed on a single degree of freedom, the case of scalar fields, for example, relies on a 
set of degrees of freedom so that $d\sigma(\pi,\phi)^2=\tint[B(x)\,d\pi(x)^2+B(x)^{-1}\,d\phi(x)^2]\;dx$, where $B(x)>0$ is a fixed positive field.

These variables enjoy $dp\wedge dq$ as measures 
of the appropriate phase space. 
The same can be said about $\tint\{d\pi(x)\wedge d\phi(x)\}\,dx$.

\section{Loop Quantum Gravity}
Using canonical quantization, the case of loop quantum gravity involves two sets of fields classically denoted by $E^a_i(x)$
and $A^i_a(x)$, where $a,i=1,2,3$, and $x$ denotes a 3-dimensional spatial point in space.
These variables admit the phase-space measure $ \tint\{dA^i_a(x)\wedge dE^a_i(x)\}\,dx$. However, their natural metric expressions, such as
$d\sigma(A,E)^2=\tint [ C(x) \,( E^a_i(x)\,dA^i_a(x))^2+C(x)^{-1}\,(A^i_a(x)\,dE^a_i(x))^2]\, dx$, where $0<C(x)<\infty$, 
fail to exhibit suitable Cartesian coordinates, and thus signal a false quantization because it does not follow Dirac's rule.

\section{Affine Quantization}
What is affine quantization? While canonical quantization employs $Q$ and $P$, with $[Q,P]=i\hbar$, as basic operators, affine quantization employs $Q$ and $D\equiv\half(PQ+QP)$, the dilation operator, with $[Q,D]=i\hbar\,Q$; note: the operator $D$ can be self-adjoint even when $Q>0$ is self-adjoint, but then $P$ can not be self-adjoint.

There are some systems that canonical quantization can solve, and there are some systems that affine quantization can solve. If they solve using one system they typically fail to solve using the other system. For example, canonical quantization can solve the Hamiltonian $H=\half( p^2+q^2)$, where $-\infty<p, q<\infty$, while affine can not solve it. On the other hand, the same Hamiltonian, $H = \half(p^2+q^2)$, now with $-\infty<p<\infty$ and $0<q<\infty$, can be solved with affine quantization but not with canonical quantization. This example is used to illustrate the power of affine quantization in \cite{JK}, and it points the way to affine quantum gravity. 

Articles \cite{JK,bqg} offer an approach to resolve quantum gravity by affine 
quantization, and they lead to positive results. 
Although paper \cite{bqg} is older, the author recommends that \cite{JK} is read first. This recommendation  is because \cite{JK} employs a familiar Schr\"odinger representation, while \cite{bqg} normally employs a less familiar current commutation representation. 

The representations of the analysis in these two papers may be different, but the physics is the same: specifically, for example, the quantum gravitational metrics are not discrete, but continuous.

\end{document}